\documentclass[11pt]{article}
\usepackage{amsmath}
\usepackage{graphics}
\usepackage[dvips]{graphicx}
\usepackage[left]{lineno}
\usepackage{multicol}

\usepackage{enumitem}
\setlist[enumerate]{itemsep=0mm}
\usepackage{caption}

\usepackage{txfonts}
\usepackage{pstricks}
\def\aj{AJ}%
\def\apjs{ApJS}%
\def\aap{A\&A}%
\def\mnras{MNRAS}%
\begin{document}

\pagestyle{myheadings}
\markright{\footnotesize {doi:\#/jcis.2019.**.**.**** \hspace{3.5cm} 
Santiago-Bautista et al.}}  

\vspace{1.2cm}

\begin{center}

{\bf  {\Large Internal structure of superclusters of galaxies from 
pattern recognition techniques}}
\bigskip


{\small Santiago-Bautista, I.$^{a, b}$;
Caretta, C.A.$^a$\footnote{E-mail 
corresponding author: caretta@astro.ugto.mx}; 
Bravo-Alfaro, H.$^a$;
Pointecouteau, E.$^b$ and
Madrigal, F.$^c$
}
\smallskip

\end{center}

{\footnotesize
\noindent $^a$ Departamento de Astronom\'ia, DCNE-CGT, Universidad de Guanajuato, 
Guanajuato, Mexico \\
\noindent $^b$ IRAP, Universit\'e de Toulouse, CNRS/CNES/UPS, 
Toulouse, France \\
\noindent $^c$ CNRS, LAAS, Toulouse, France
}


\quad


\begin{abstract}
The Large-Scale Structure (LSS) of the Universe is a homogeneous network 
of galaxies separated in dense complexes, the \emph{superclusters of galaxies}, 
and almost empty \emph{voids}. The superclusters are young structures that did 
not have time to evolve into dynamically relaxed systems through the age of 
the Universe. Internally, they are very irregular, with dense cores, 
filaments and peripheral systems of galaxies. We propose a methodology to 
map the internal structure of superclusters of galaxies using pattern 
recognition techniques. Our approach allows to: i) identify groups and 
clusters in the LSS distribution of galaxies; ii) correct for the 
``fingers of God'' projection effect, caused by the partial knowledge of 
the third space coordinate; iii) detect filaments of galaxies and trace 
their skeletons. In this paper, we present the algorithms, discuss the
optimization of the free parameters and evaluate the results of its
application.
With this methodology, we have mapped the internal structure of 42 
superclusters in the nearby universe (up to $z=0.15$).

\quad

\end{abstract}


\quad


\textbf{1. Introduction \label{Intro}}
\smallskip

One important problem in Cosmology is related to the description and
mapping of the Large Scale Structure of the Universe (LSS). 
It has passed about half a century since we had the first clear 
glimpse of galaxies not being randomly distributed, but organized in a 
network-like pattern composed by groups, clusters, filaments and walls, 
separated 
by voids (e.g., \cite{GH1989}).
The regions where the galaxy flux (peculiar velocities) tends to 
concentrate the matter are called \emph{superclusters of galaxies}, 
while the \emph{voids} are outflow regions.
Thus, the superclusters can be described as a network of elongated structures 
(filaments of galaxies), crossing each other in denser knots (clusters 
and groups of galaxies). 

In the current cosmological standard model, 
dark matter 
is the main driver for gravity and, as a consequence, for the formation 
and evolution of LSS, 
while the baryonic matter, 
gas and galaxies, populate this structure accordingly.
Nevertheless, the formation of this network cannot be completely 
described by current Physics theories. 
Nowadays, the most accepted model is still the 70's Zel'dovich
approximation (\cite{Zeldovich1970}, see also the ``sticking model'' by 
\cite{Shan-Zeld1989}), although it concerns only to the beginning of the 
non-linear regime of cosmic evolution.

In order to identify, map and study the superclusters and the LSS
in general, it is necessary to make use of mathematical and computational 
tools suitable for detecting and describing such complex structures 
and topologies.
In this work, we present a methodology that uses clusterization and 
graph methods to identify both compact and relatively dense regions 
(groups and clusters of galaxies) and lower density elongated structures 
(filaments), from 3D galaxy data.

The present work is organized as follows.
In Sec. 2,
we describe the astrophysical case that we have
to take into account as background and boundary conditions. In Sec. 3,
we present the first part of the method, dedicated to
find the galaxy systems and correct the data for the ``fingers of God'' 
effect, while in Sec. 4
we present the second part,
devoted to find the filaments. Sec. 5
presents a
discussion of our results and some conclusions about the
application of the method and its evaluation.

\quad
\bigskip

\noindent \textbf{2. Astrophysical aspects \label{Astro}}
\smallskip

\textit{2.1 Astronomical coordinates}

In Astronomy, the position of a celestial object is set by its sky projected
coordinates, $\alpha$ and $\delta$ (respectively, right ascension and 
declination), and by an estimate of its distance. The distance is one of 
the most important and, at the same time, difficult parameters to measure.
Beyond a certain distance, we use the spectroscopic redshift of the galaxies, 
$z$, to measure their radial velocity, which is proportional to their 
distance in the absence of peculiar motions. 
Thus, we can obtain an approximate rectangular coordinate system from:
\vspace{-25pt}
\begin{center}
   \begin{align}
	X&=D_C~\cos\left(\delta \right) \cos\left( \alpha \right), \label{transformation1}\\
	Y&=D_C~\cos\left(\delta \right) \sin\left( \alpha \right), \label{transformation2}\\
	Z&=D_C~\sin\left(\delta \right),\label{transformation3} 
   \end{align}
\end{center}

\noindent where $D_C$ is the co-moving distance obtained by using the 
redshift and the cosmological parameters:
\vspace{-25pt}
\begin{center}
   \begin{align}
	D_C &= \frac{c}{H_0} \int_{0}^{z} \frac{dz'}{E(z')}, \\
	E(z) = \frac{H(z)}{H_0} &= \sqrt{\Omega_r (1+z)^4 + \Omega_m (1+z)^3 + \Omega_k (1+z)^2 + \Omega_{\Lambda}}.
   \end{align}
\end{center}

\noindent Throughout this paper  we assume the Hubble constant 
$H_0 = 70~h_{70}$~km~s$^{-1}$~Mpc$^{-1}$, 
the matter density $\Omega_m=0.3$, 
the dark energy density $\Omega_\Lambda=0.7$,
and the radiation ($\Omega_r$) and curvature ($\Omega_k$) 
densities very close to 0.

Nevertheless, we know galaxies might have peculiar velocities, especially 
if they are members of clusters and groups. 
This is called the ``fingers of God'' (FoG) 
effect, a stretch of the redshifts along the line-of-sight of these galaxy 
systems caused by their velocity dispersion. 
Since this effect is critical for the kind of study we are doing, one part 
of our strategy is devoted just to correct this effect.

\bigskip
\quad
\textit{2.2 Galaxy data}

Since we are interested in mapping the LSS through the distribution of the
galaxies, we used one of the largest galaxy databases available to the
moment, the Sloan Digital Sky Survey (SDSS, DR-13, \cite{DR13}).
This database covers more than one third of the sky (14,555 square degrees) 
and contains optical photometry, positions (with an astrometric precision of 
0.1 arcsec) and spectroscopic redshifts for about 3 million galaxies.

We extracted SDSS galaxies inside boxes that contain superclusters of 
galaxies, the last ones selected from the Main SuperCluster Catalog 
(MSCC, \cite{chow2014}). This all-sky superclusters catalog is limited to
redshift $z = 0.15$, while 45 of its superclusters are completely contained 
inside the SDSS footprints. The SDSS sampling is sparse, but dense enough for 
the goals of the present study.

In the following, we considered the $N$ galaxies in each supercluster box 
volume as a set of points ${x_1,x_2,...,x_N} \in X$, all being part of a 
sample $X$.

\bigskip
\quad
\textit{2.3 Density of galaxies}

We used Voronoi tessellation (VT, \cite{Voronoi1908} to calculate the local 
density at each galaxy position, both in 2D (projected densities) and 3D 
(volume densities).
VT is a well-known technique (e.g., \cite{Scoville2013, Darvish2015}),
which partitions the space into optimal polygonal cells in a way that there 
is one cell for each galaxy position ${x_i} \in X$.
Then,  the density at  $x_i$ is determined as $d_i=1/v_i$, with $v_i$ being 
the volume (or area) of the cell around the galaxy $x_i$.

\bigskip
\quad
\textit{2.4 Virial mass estimate}

Another highly important (and difficult) measurement in Astrophysics 
is the mass of an object or system. Together with the distance, these are
the parameters that define gravity and, thus, evolution.
For our work we are interested in estimating the mass of the galaxy systems
in order to correct the positions of their member galaxies for the FoG effect.

We used a simplified version of the algorithm proposed by \cite{Biviano2006}
to iterate the virial masses (under the assumption of dynamical equilibrium,
relative isolation of the systems and roughly spherical shapes). 
In summary, this algorithm works as follows:
i) The galaxies are selected, in the $\alpha\times\delta$ projection,
inside a cylinder of radius $R_{a} = 1~h_{70}^{-1}$~Mpc, with a length
in the line-of-sight direction of  
${\Delta}z = 0.02$ ($\pm 3,000$~km~s$^{-1}$), centered at the brightest
galaxy of the system and at a previously estimated mean velocity;
ii) A robust estimation of mean velocity, $v_{LOS}$, and velocity 
dispersion, $\sigma_v$, for the galaxies inside the cylinder, is 
calculated by using Tukey's biweight (See eq. 9 in \cite{Beers1990}); 
iii) The mass inside the cylinder is estimated as: 
$M_a = (3\pi/2G)~\sigma_{v}^{2}~R_{h}$, 
where G is the gravitational constant, 3$\pi$/2 is the deprojection 
factor and $R_{h}$ is the projected harmonic radius;
iv) The virial radius is calculated assuming a spherical model for 
nonlinear collapse as:
 \begin{equation}
\label{R_V}
R_{vir}^3 = \frac{3}{4\pi} \frac{M_{vir}}{\rho_{vir}} = \frac{\sigma_v^2~R_{h}}{6\pi~H(z)^2}. 
 \end{equation}
\noindent using $M_{a}$ as an estimation for $M_{vir}$ and, for the 
virialization density, $\rho_{vir} = 18\pi^2 [3 H^2(z)] / [8 \pi G]$;
v) Then, the aperture $R_a$ is updated to the calculated $R_{vir}$ value,
the mean velocity to $v_{LOS}$ and the ${\Delta}z$ to the $\pm 3 \times 
\sigma_v$. 

The steps i-v are repeated iteratively until the $R_{vir}$ value 
converges. In the end, we calculate $M_{vir}$ from equation \ref{R_V}.
The convergence takes, typically, six iterations.

\quad
\bigskip

\noindent \textbf{3. Galaxy systems finding algorithm(GSyF) \label{Systems}}
\smallskip

In order to identify clusters and groups of galaxies (which we refer
generically, along this text, as galaxy systems),   
we developed an algorithm that detects such systems and allows to correct 
the position of their galaxy members for the FoG effect.
The implemented methodology consists on:
i) First we calculate the local surface density for each galaxy in the
supercluster volume by applying the VT technique to their projected 
distribution in the sky plane, in this case in units of deg$^{-2}$.
ii) From these densities, we calculate the density contrast, $\delta_i$, 
respect to a baseline (background) density value, $d_{bas}$, estimated by 
simulating 1,000 random isotropic galaxy distributions with the same contour 
conditions:

\begin{equation}
d_{bas}=\frac{1}{m} \sum_{j=1}^{m}\frac{1}{n} \sum_{i=1}^{n}d'_{i,j},
\end{equation}
\noindent where $d'_{i,j}=1/v'_{i,j}$ corresponds to the area of the point 
$x'_i$ for the randomization $j$, and:
\begin{equation}
\delta_i=\frac{d_i-d_{bas}}{d_{bas}}.
\end{equation}

\noindent iii) We consider only the $N_{gal}$ galaxies with density 
contrasts above certain value ($\delta_i \geq g$ or, equivalently, 
$d_i \geq (1+g)~d_{bas}$) for detecting the systems. 
iv) Then, we apply a hierarchical clustering (HC) algorithm to group the 
galaxies by their position ($\alpha, \delta$ and $z$, with a weight of 
1,000 for $z$). For this work we used an agglomerative HC method and the 
Ward's minimum variance clusterization criteria, described in detail 
in \cite{Murtagh2014}. 
The number of HC groups to be extracted is fixed as 
$N_{groups} = N_{gal}/f$, with a segmentation parameter $f$ which is the 
expected mean number of elements per group. Only groups with $N_i \geq 3$
member galaxies are retained.
v) After, we use the mean projected position and the mean redshift of the 
HC group to calculate its virial radius and mass by using the strategy 
described in the previous section. 
This process allows to refine the identification of the groups, 
both by eliminating unlikely real systems and by re-grouping 
duplicated/over-clustered groups. 
vi) Finally, the membership of the systems, $N_{mem}$, has been established 
and we can correct the positions of these galaxies by scaling their 
comoving distances along the line-of-sight to the calculated virial
radius.
The parameters $f$ and $g$ were set by an optimization test, described in 
the next subsection.

\bigskip
\quad
\textit{3.1 Optimization of GSyF parameters}

The detection efficiency of the GSyF method may depend significantly on the 
election of $f$ and $g$ parameters.
In order to find the best choice for them, we constructed a set of 30 mock
simulations for each supercluster of the sample.
We used the relations estimated by \cite{Pearson2015} to determine the  
properties of the synthetic galaxy systems.
These galaxy-based relations were calculated for the SDSS database and 
follow a power law of the form 
$\log_{10}(M_{500})=\alpha\log_{10}(\frac{x}{x_0})+\beta$, where $x$ is
the property under consideration.

Each mock map is constructed as follows:
the simulated volume is filled with $N_{synth}$ synthetic systems, randomly
distributed, each with synthetic galaxies in the range $N_{elem} = {10-200}$.
The number of systems in the volume is set using the power function 
\ $\log_{10}[N_{synth}(N_{elem})] [h_{70}^3$~Mpc$^{-3}] = m\log_{10}(N_{elem}) + b$ \ 
(hereafter, multiplicity function \cite{Berlind2006}).
The slope and intercept ($m$, $b$) are set as ($-2.48$, $-2.1$) for $z<0.08$
and ($-2.72$, $-2.4$) for higher $z$ values.
The system proxies ($M_{500}$ mass, $R_{vir}$ radius and $\sigma_v$ 
velocity dispersion) are calculated from: 
\vspace{-25pt}
\begin{center}
   \begin{align}
      \log_{10}M_{500}&=1.03\log_{10}(N_{elem}-2.63)+0.34 \\
      \log_{10} R_{vir}&=1.05~\log_{10}(M_{500}-0.35)+8.48 \\
      \log_{10} \sigma_v^3&=2.33~\log_{10}(M_{500}-0.21)+3.04	
   \end{align}
\end{center}

\noindent which were obtained by least-square fitting (BCES estimator)
by \cite{Pearson2015}.
Therefore, the systems are filled with galaxies following a normal 
distribution $N_{elem}({\mu_i}, R_{vir})$ with $\mu_i = \{\alpha_i, \delta_i\}$.
Then, the FoG effect is incorporated to the system' galaxies by 
adding a velocity dispersion $N_{elem}(z_i, \sigma_v)$.
Finally, random galaxies are added to the box volume following the ratio: 
60\% of galaxies are distributed in the field and 40\% in the systems.

\begin{figure}[h!]
   \includegraphics[width=5cm]{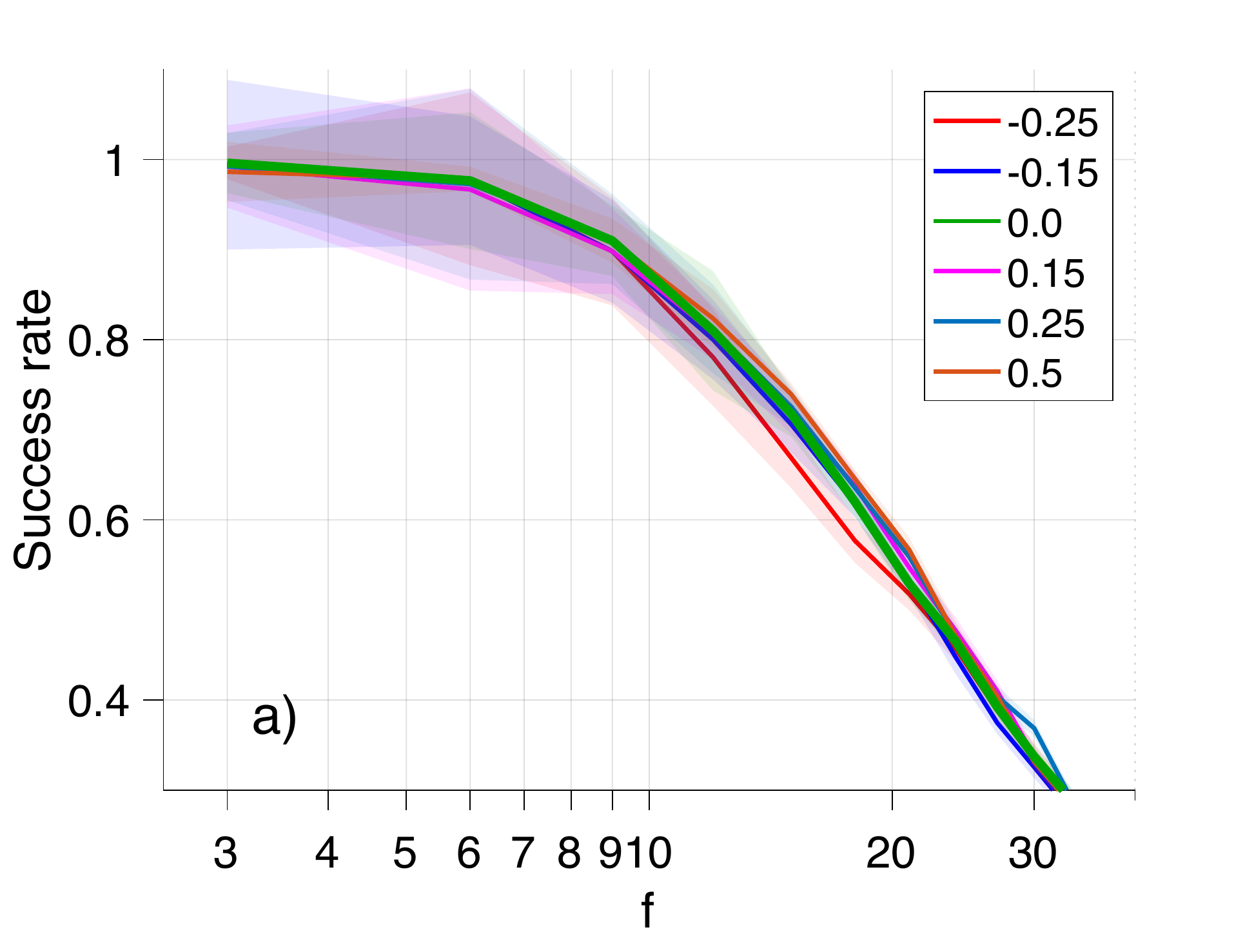}
   \includegraphics[width=5cm]{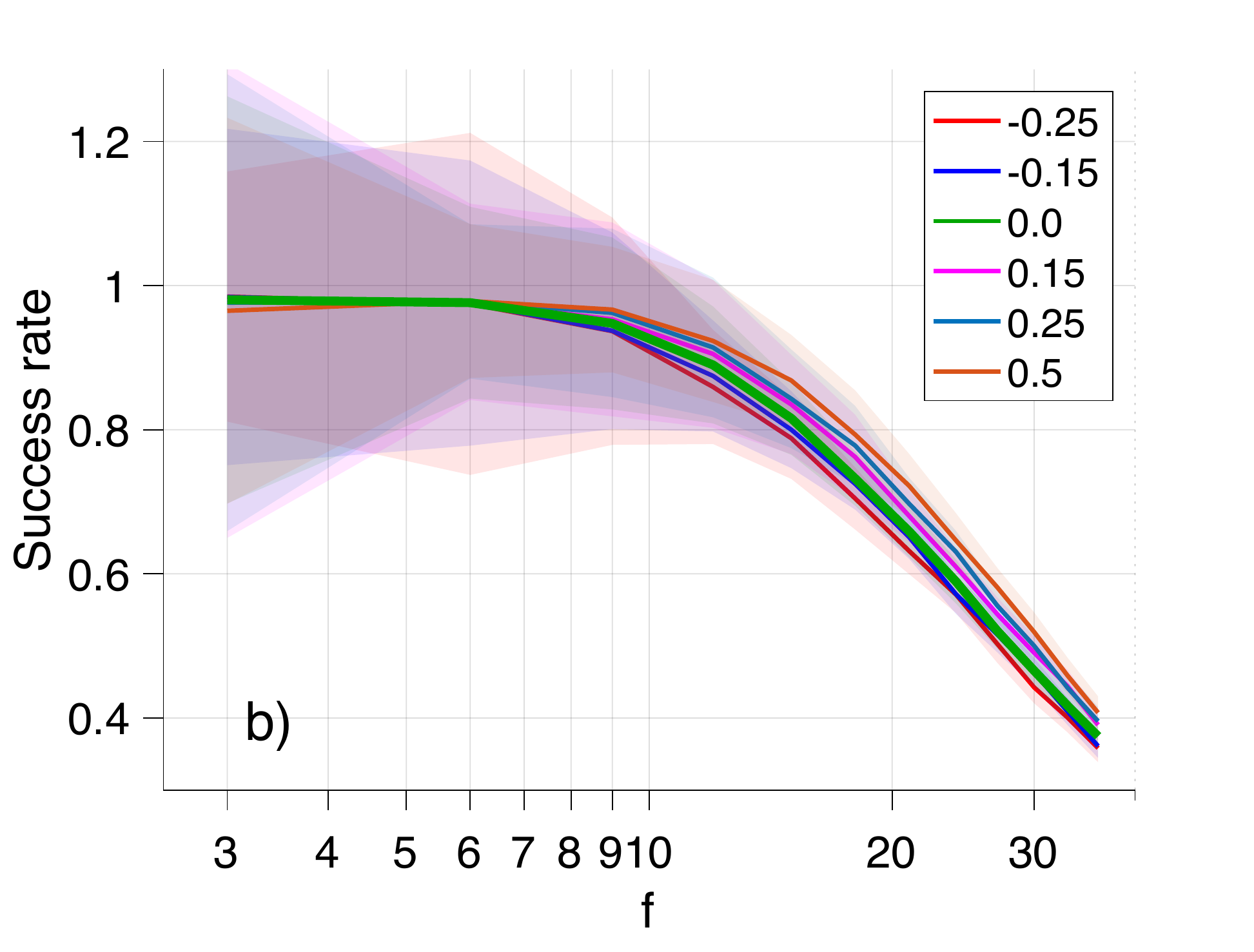}\\
   \includegraphics[width=5cm]{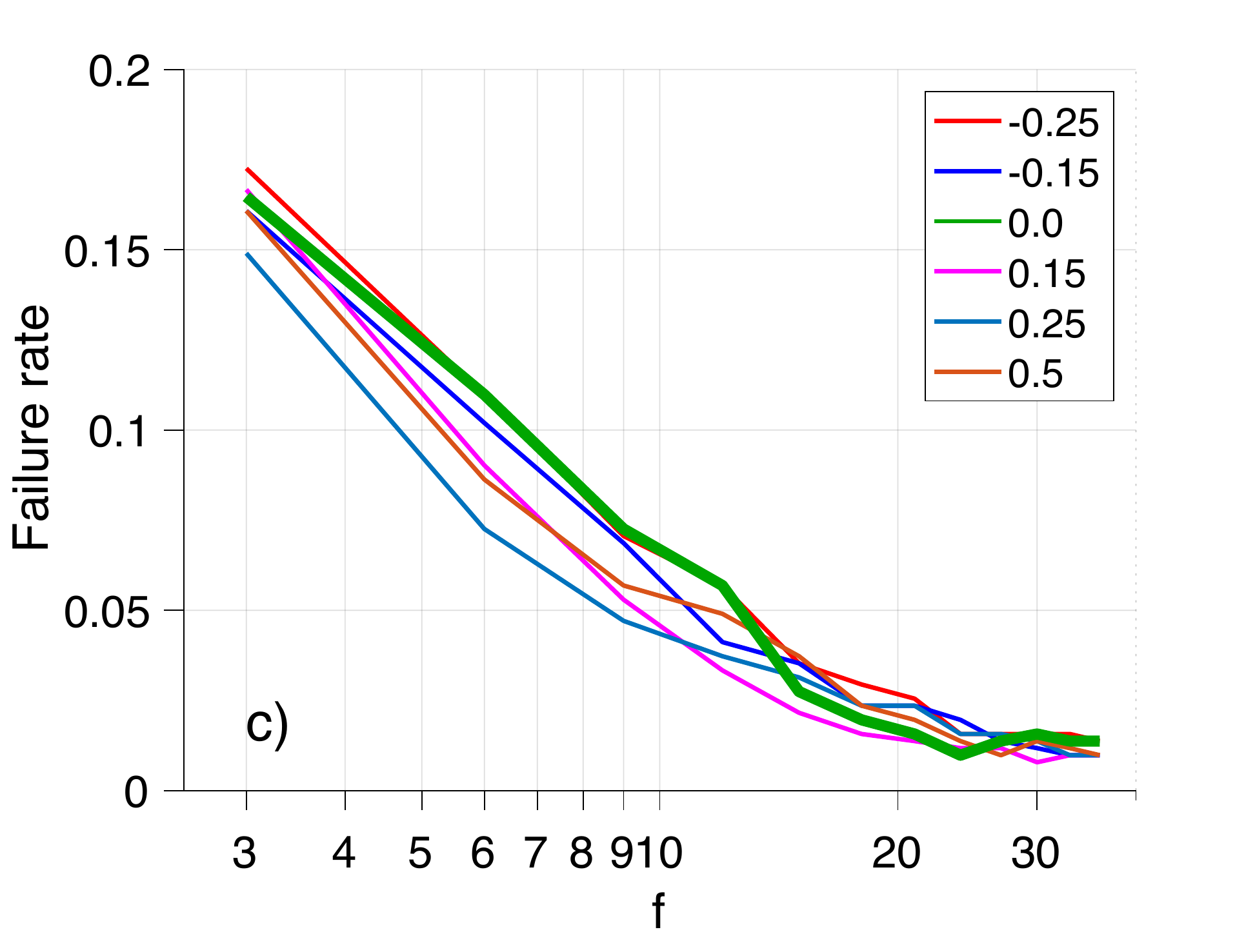}
   \includegraphics[width=5cm]{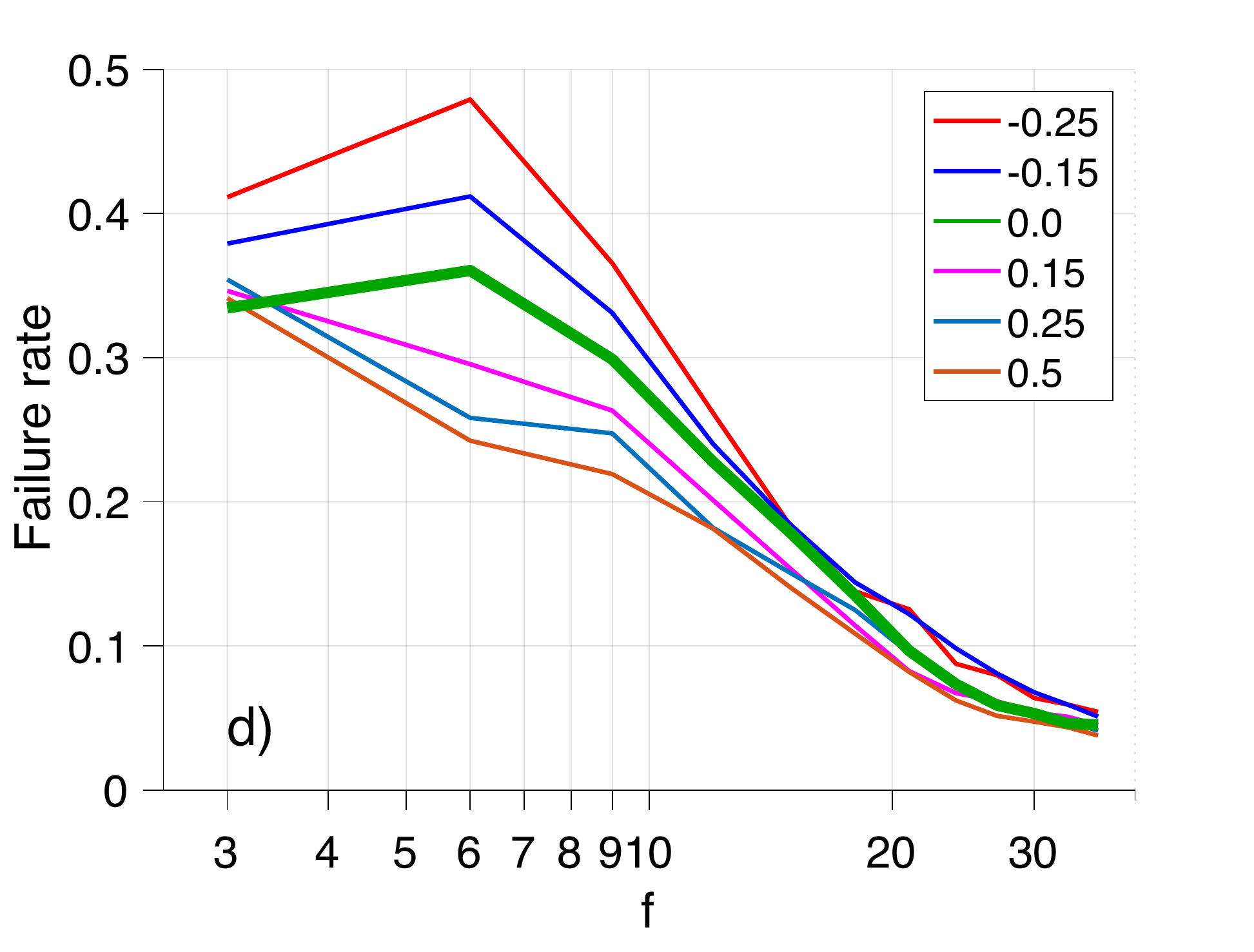}\\
   \includegraphics[width=5cm]{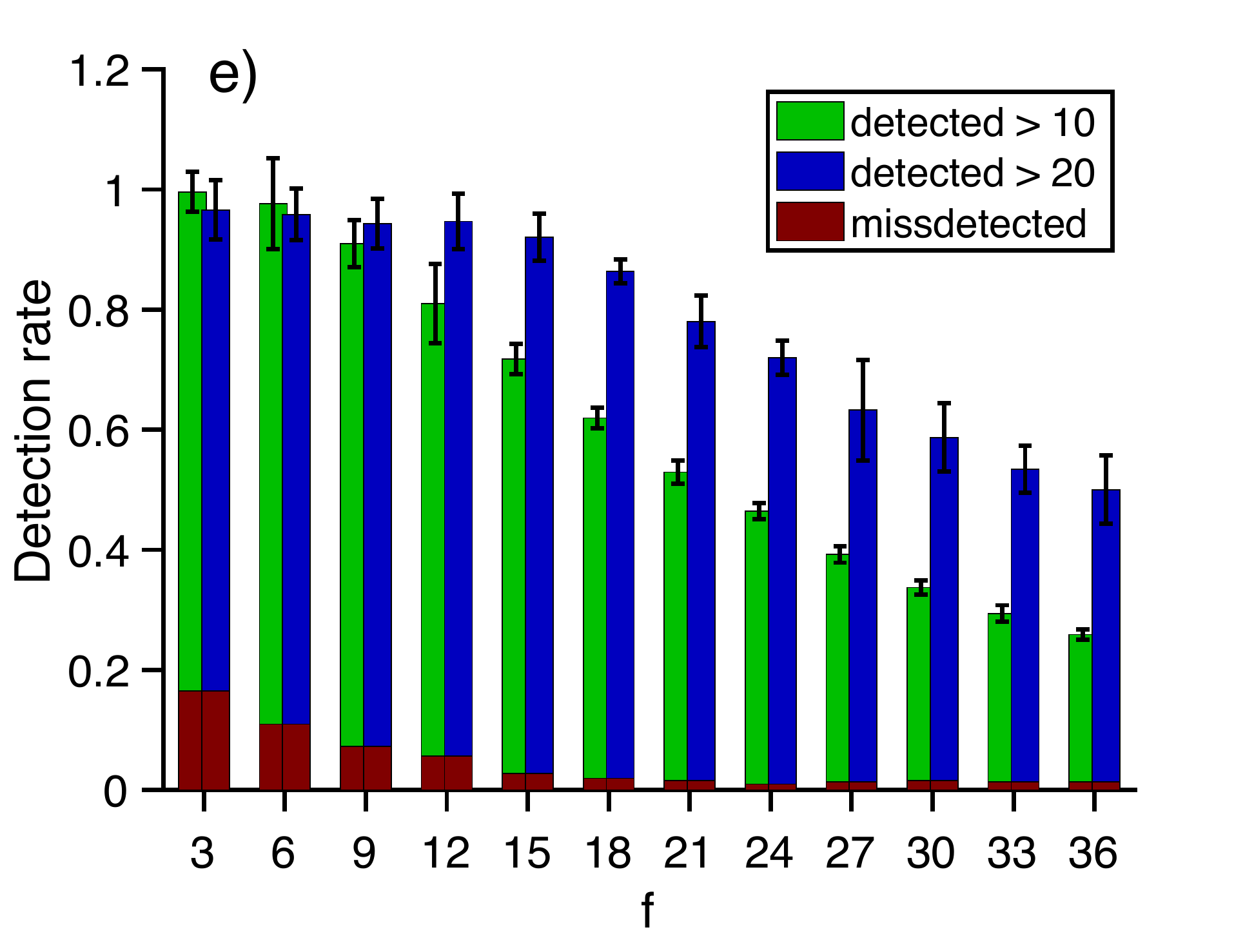}
   \includegraphics[width=5cm]{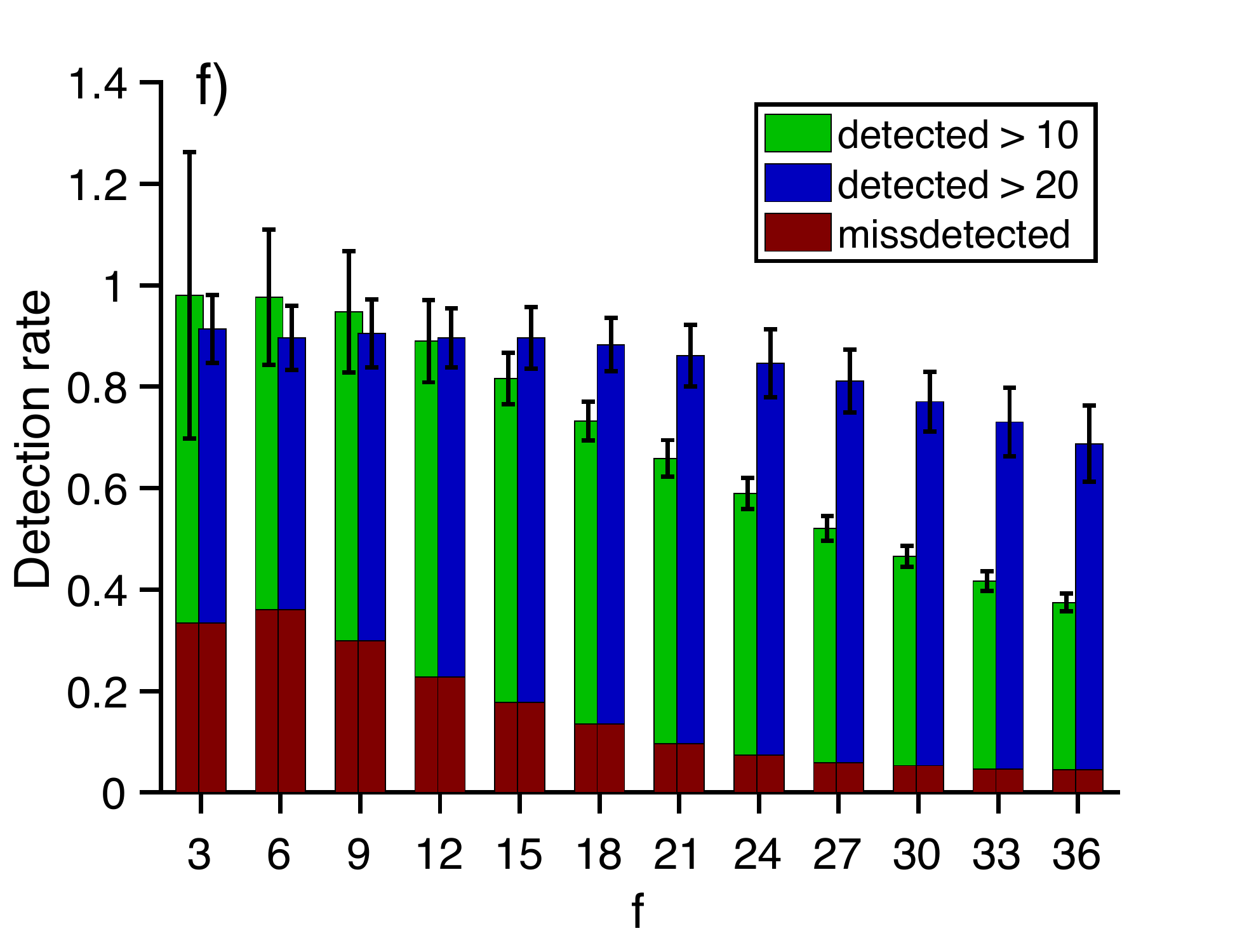}\\
      \caption{\footnotesize Optimization of the segmentation, $f$ 
($x$ axis), and minimum contrast density, $g$ (lines of different colors
on panels a--d), for the 
supercluster volumes of MSCC-310 ($z$=0.06) and MSCC-454 ($z$=0.04), 
from the mock simulations. (a, b) Success rates; (c, d) failure rates; 
(e, f) detection rates for systems with $N_{mem} \geq 10$ (green) and 
$N_{mem} \geq 20$ (blue), with $N_{mem}$ been the final number of 
members after the virial refinement. The failure rates are shown in 
brown. \label{fig:optmGSyF}}
\end{figure}

We applied the GSyF method over the mock maps probing values for $f$
in the range $f=\{3, 6, 9,..., 36\}$ and for $g$ in the range 
$\{-0.25, -0.15, 0.0, 0.15, 0.25, 0.5\}$.
The first range corresponds to the typical number of galaxies in a 
group expected for the kind of data we have, while the second explores 
the values of density contrast around zero.
Our results suggest that $g$ does not have a significant impact on the 
efficiency of the algorithm, as can be seen in Fig. \ref{fig:optmGSyF}
(panels a--d).
Therefore, we adopted the value of $g = 0.0$ for all superclusters.

For the $f$ parameter, on the other hand, we can see from Fig. 
\ref{fig:optmGSyF} (same panels) that small values (below 10) allow higher 
completeness (above 85\%, arriving close to 100\% for $f <$ 5), while 
the contamination, in the same range, is critically between
5 and 35\%, getting better for larger $f$.
Fig. \ref{fig:optmGSyF} (panels e--f) reveals that the efficiency of the 
algorithm also depends on the final richness of the systems: success 
rates for richer systems ($N_{mem} \geq 20$) are much higher than success 
rates for poor ones ($N_{mem} \geq 10$), while failure rates do not change 
significantly. 
We select optimal values for $f$ as 
those that maximize 
the detection of synthetic systems (completeness) and minimize the number of 
false detections (contamination), that is, the values that maximize the
function:

\begin{equation}
\Gamma = \frac{N_{detec}}{N_{synth}} +\left( 1 - \frac{N_{fail}}{N_{synth}}\right) .
\end{equation}

\begin{figure}[t!]
	\includegraphics[width=4cm]{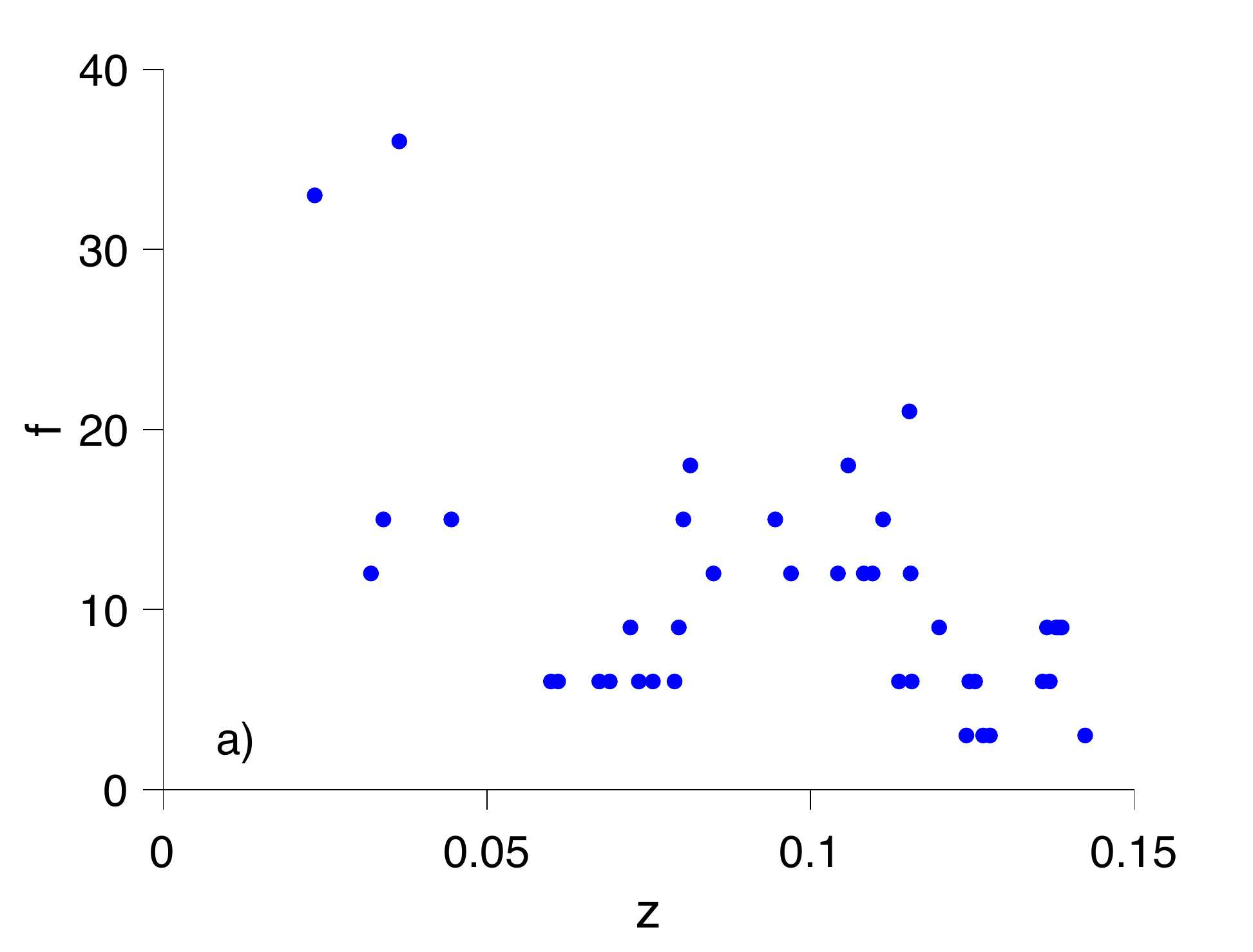}
	\includegraphics[width=4cm]{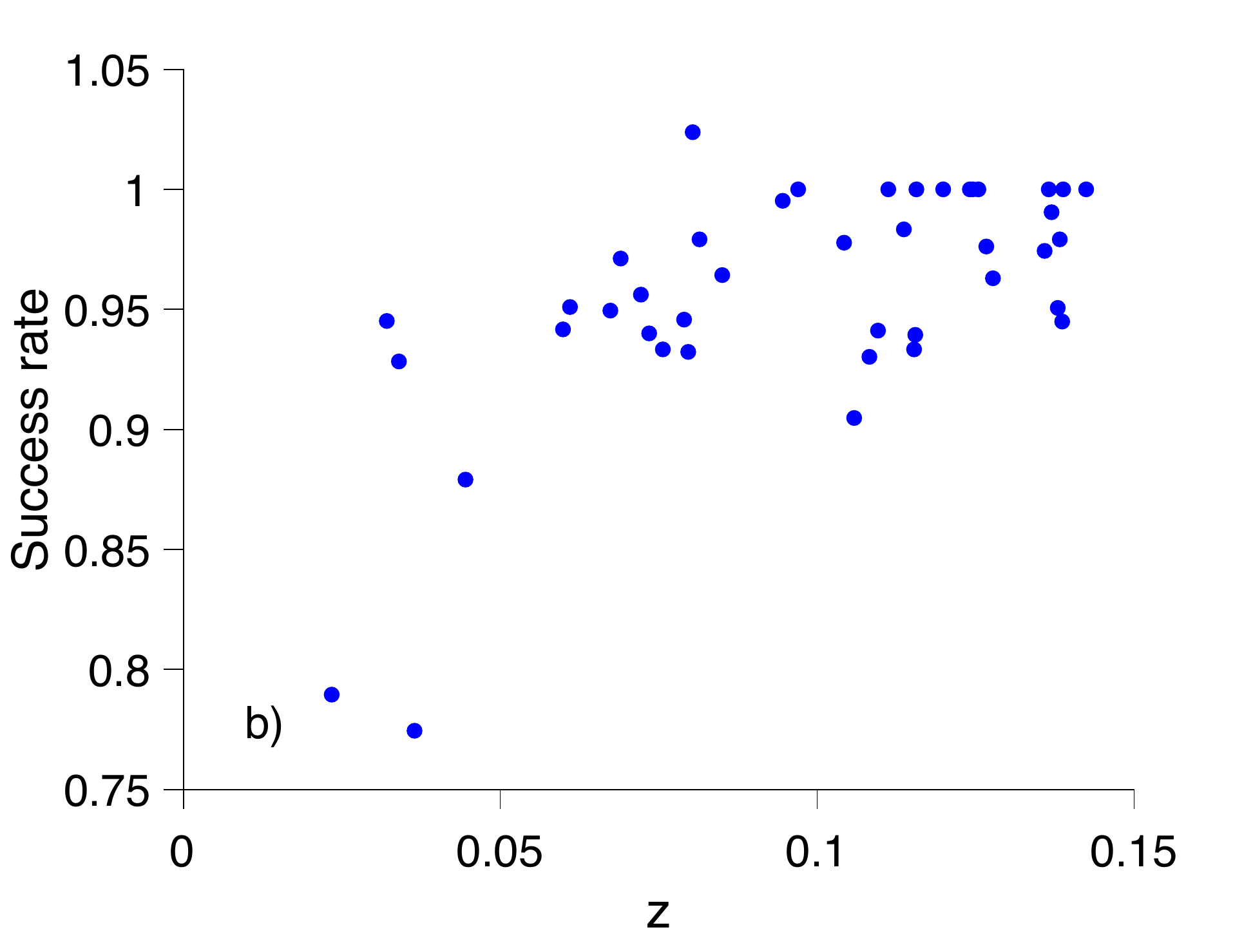}
	\includegraphics[width=4cm]{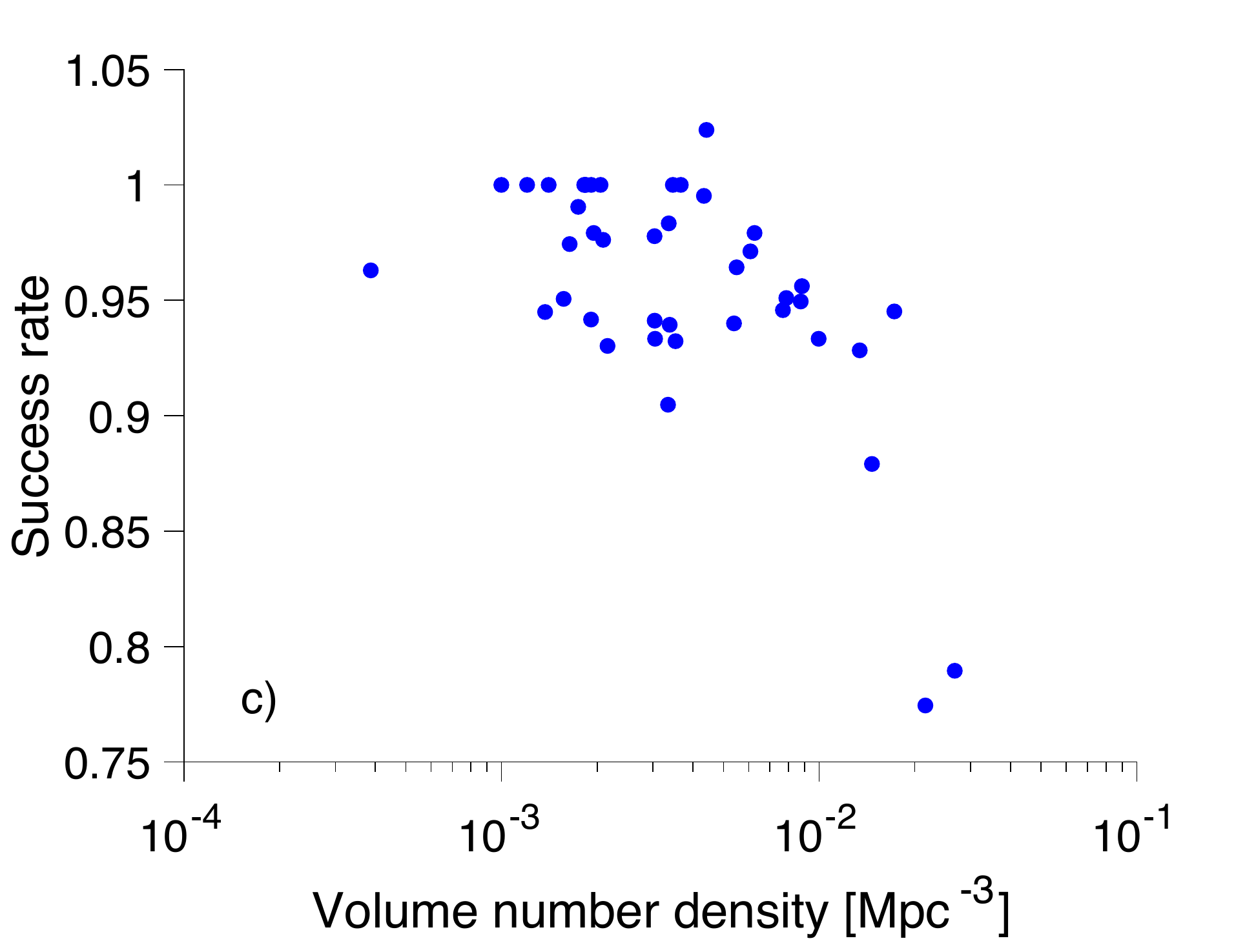}
	\caption{\footnotesize a) Distribution of best $f$ values with 
redshift ($z$).
b) Distribution of mean success rates with $z$. 
c) Distribution of mean success rates with volume number densities in 
each supercluster volume. 
Mean numbers are from the 30 mock simulations for each of the 42 
superclusters. \label{fig:fzGSyF}}
\end{figure}

Figure \ref{fig:fzGSyF}a shows the distribution of best $f$ values
for the superclusters of our sample with their mean redshift. 
Except for the first two points (the most nearby superclusters),
one can see that there is no clear correlation between the two
parameters, presenting a mean value around $f = 10$ with a considerable 
dispersion.
In fact, the two most nearby superclusters are dropping the global success 
rate because, as one can see in Fig. \ref{fig:fzGSyF}b, these superclusters
have lower success rates of $75-80\%$, while all others are around 
or above 90\% (with a mean failure rate of about 10\%). 
The low success rates for 
these two superclusters seem to be related to their higher number densities
respect to the others, as can be seen in Fig. \ref{fig:fzGSyF}c.

\quad
\bigskip

\noindent \textbf{4. Galaxy filaments finding algorithm (GFiF) \label{Filaments}}
\smallskip    

The analysis and detection of elongated structures (filaments) is 
applied to the same data we used before, but with the galaxies' positions
corrected for the FoG effect.
Initially, we detect groups almost in the same way we did with GSyF.
However, this time we use the rectangular coordinates ($X, Y, Z$) of the 
galaxies, their VT local volume densities, and do not apply the threshold
(minimum $\delta_i$) to the galaxy sample. That is, now $N_{gal} = N$.
In addition, we apply again an agglomerative hierarchical clustering HC 
using Ward's criteria to group the galaxies.

Once the list of HC groups has been obtained, we perform the GFiF 
algorithm as follows:
i) First, we measure the Euclidean distance, $D_E$, of each group centroid 
against all its neighboring groups, as well as the Bhattacharyya coefficient 
$BC$ \cite{DB1943} for the pair, in order to take into account the relative 
orientations among them. 
ii) After, we connect all the groups that have $D_E$ smaller than a threshold 
(hereafter linking length), $D_{min}$. 
The ensemble of connections is considered as an undirected graph 
$G=(U,E)$, where $U$ represents the nodes (the group's 
centroids), $E$ the edges (connections between nodes) and $W$ a 
matrix that gives weights to the connections. 
For the present analysis, we used the $BC$ coefficients as the weighting 
values in $W$.
iii) Then, we apply the Kruskal's minimum spanning tree 
algorithm (MST, \cite{Graham1985}) to extract the dominant connections 
(i.e., trees) from the general graph (see also \cite{Cybulski2014}).
We considered only trees with at least three nodes connected and located 
farther than a given threshold from the box boundaries. 
The boundary condition was applied with a platikurtic Gaussian function
included in the weightings of the MST.

\bigskip
\quad
\textit{4.1 Optimization of GFiF parameters}

We also performed optimization tests for the GFiF parameters, in a 
similar way we did for GSyF. Here we probed the linking length $D_{min}$
and the segmentation $f$ for the HC algorithm.
We applied the GFiF method over the real supercluster boxes 
testing different values of $f$ in the range  $\{8,..., 40\}$, 
and  $D_{min}$ in the range $\{3,..., 14\}~h_{70}^{-1}$~Mpc.

\begin{figure}[h!]
   \includegraphics[width=4.5cm]{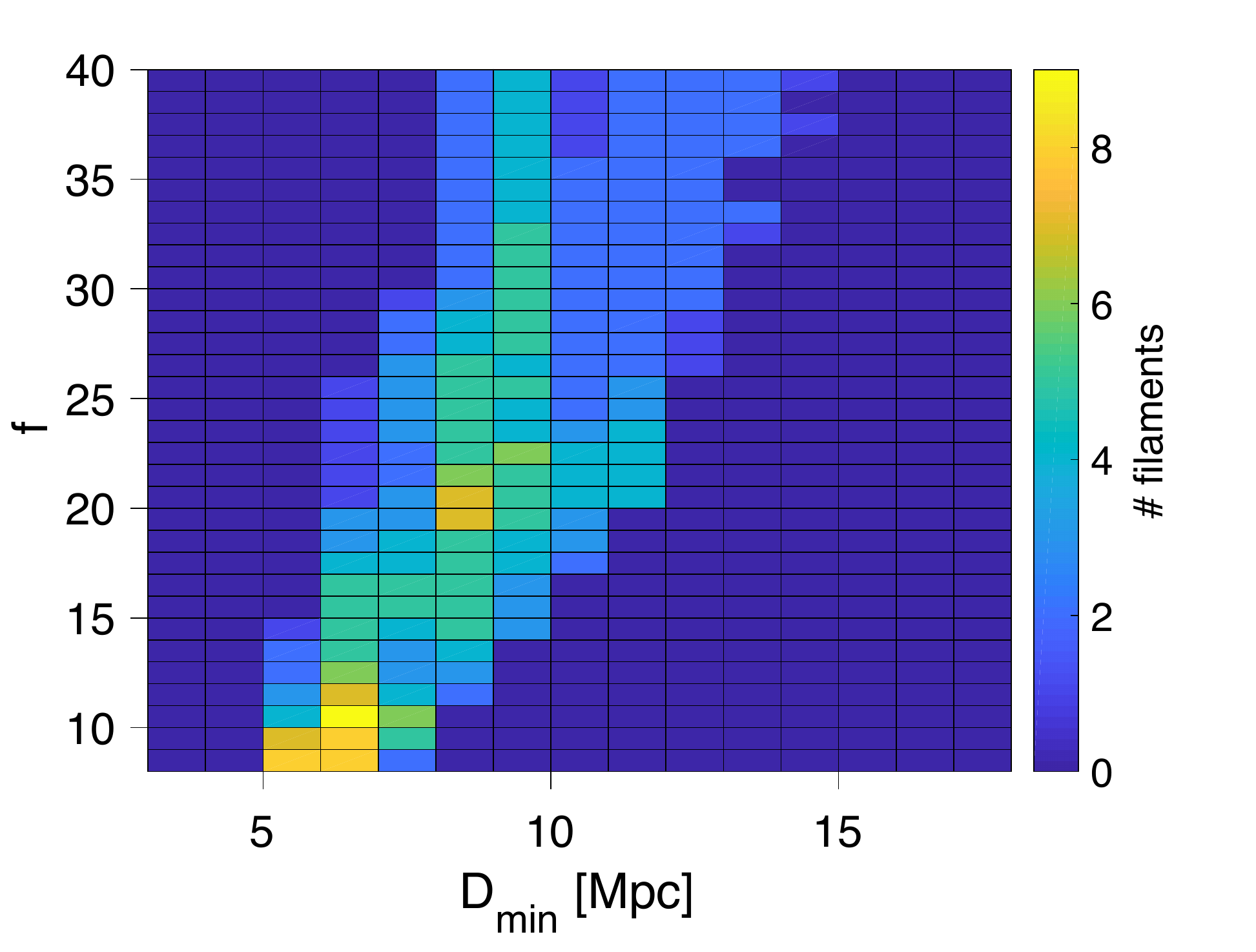}
   \includegraphics[width=4cm]{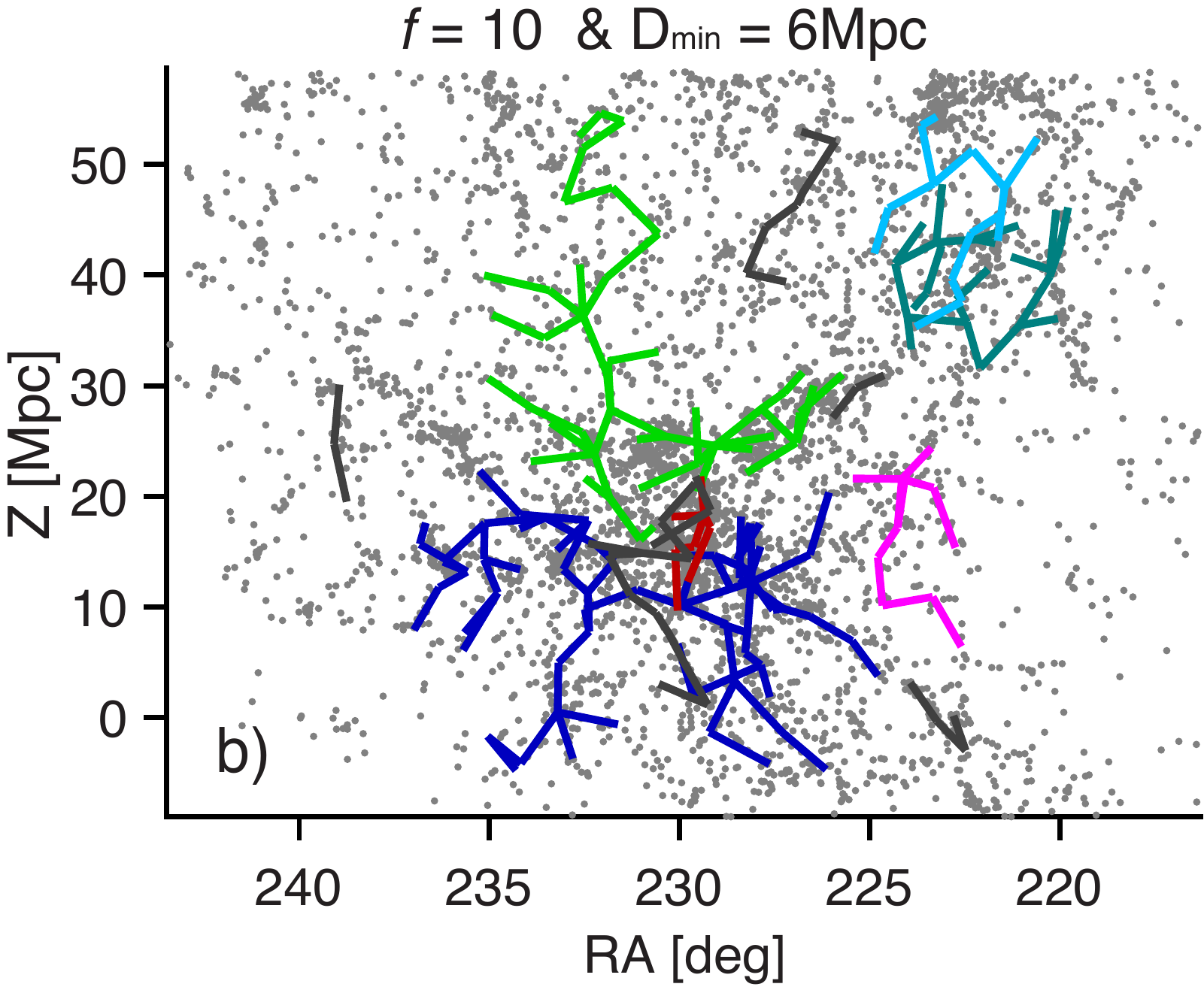}
   \includegraphics[width=4cm]{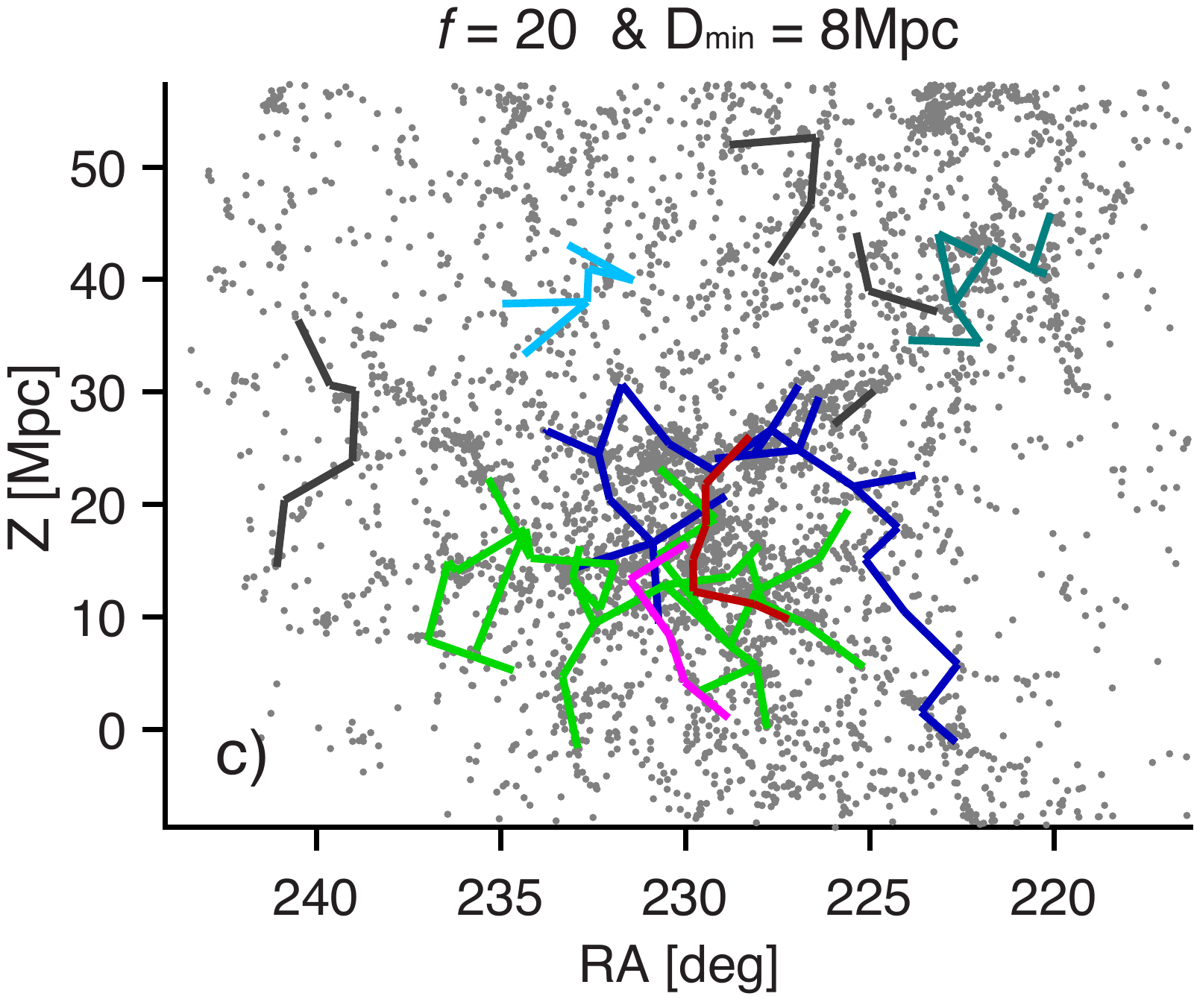}
      \caption{\footnotesize a) Optimization of the segmentation, $f$, and 
linking length, $D_{min}$, parameters for the supercluster MSCC-454. 
This optimization is based in the number of relatively dense and long 
filaments detected. 
b) Filaments found by the GFiF algorithm for the solution parameters 
$f=10~h_{70}^{-1}$~Mpc and $D_{min}=6~h_{70}^{-1}$~Mpc. 
c) Filaments found by the GFiF algorithm for the solution parameters 
$f=20~h_{70}^{-1}$~Mpc and $D_{min}=8~h_{70}^{-1}$~Mpc. }
      \label{Fil_test}
\end{figure}

\begin{figure}[h!]
   \includegraphics[width=4cm]{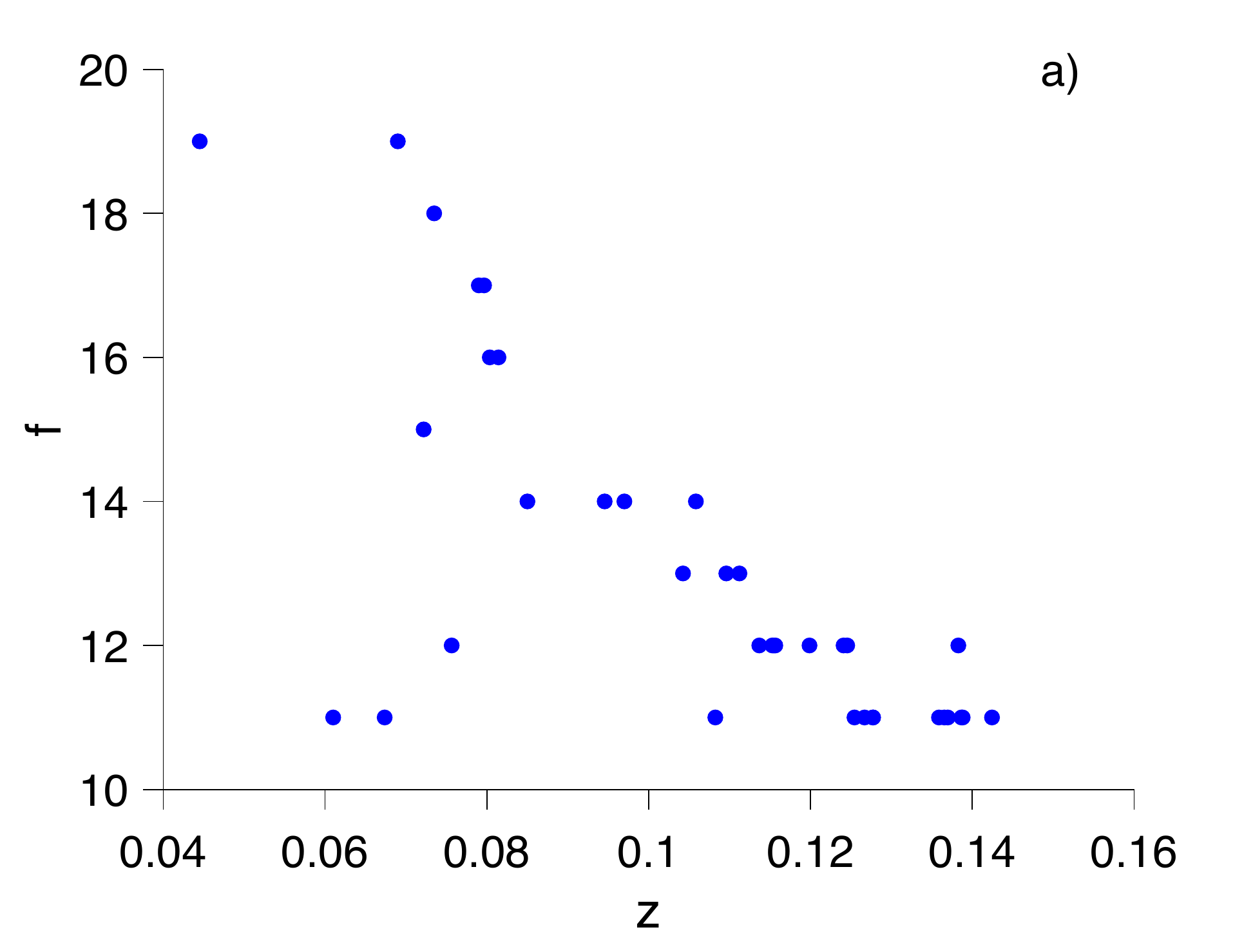}
   \includegraphics[width=4cm]{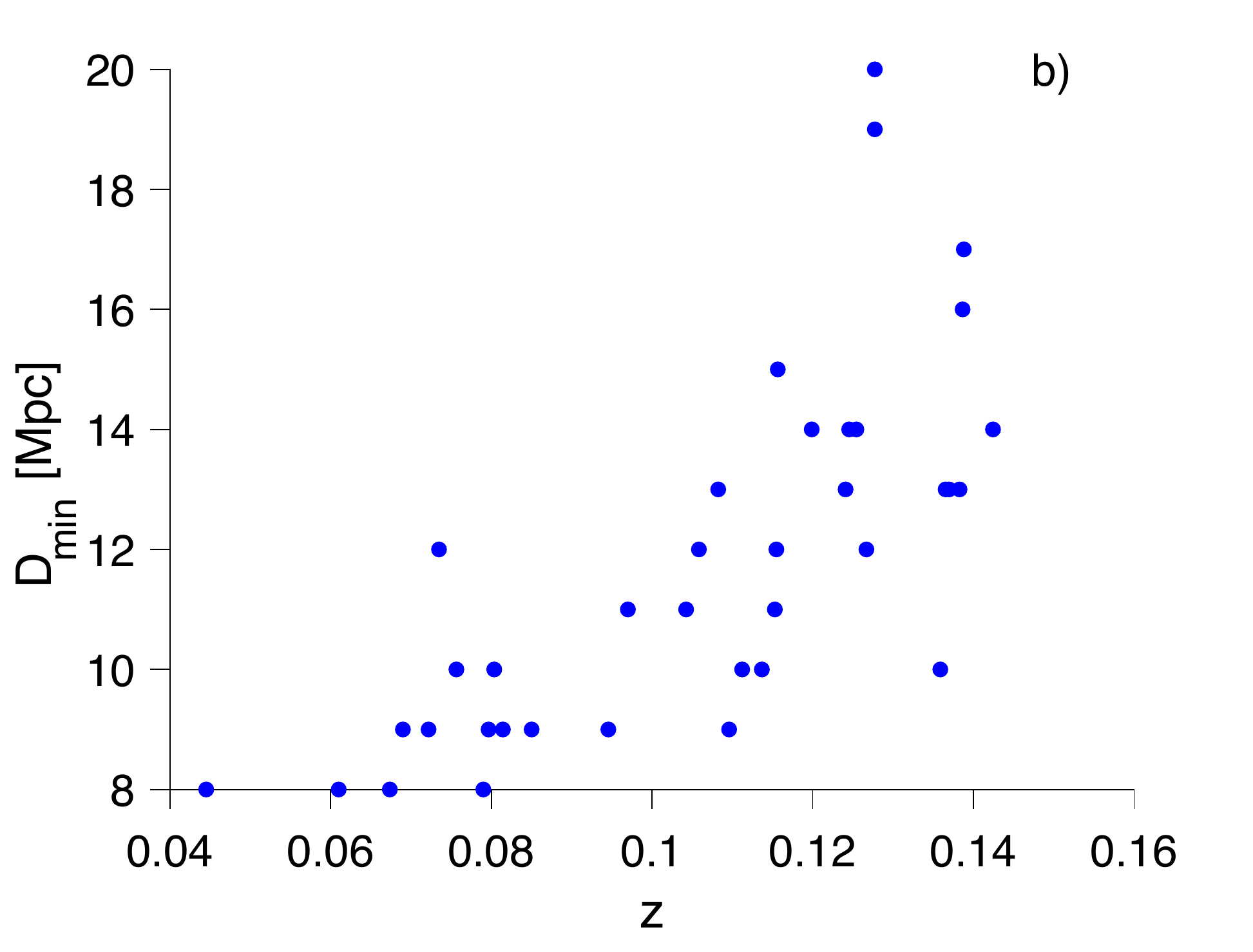}
   \includegraphics[width=4cm]{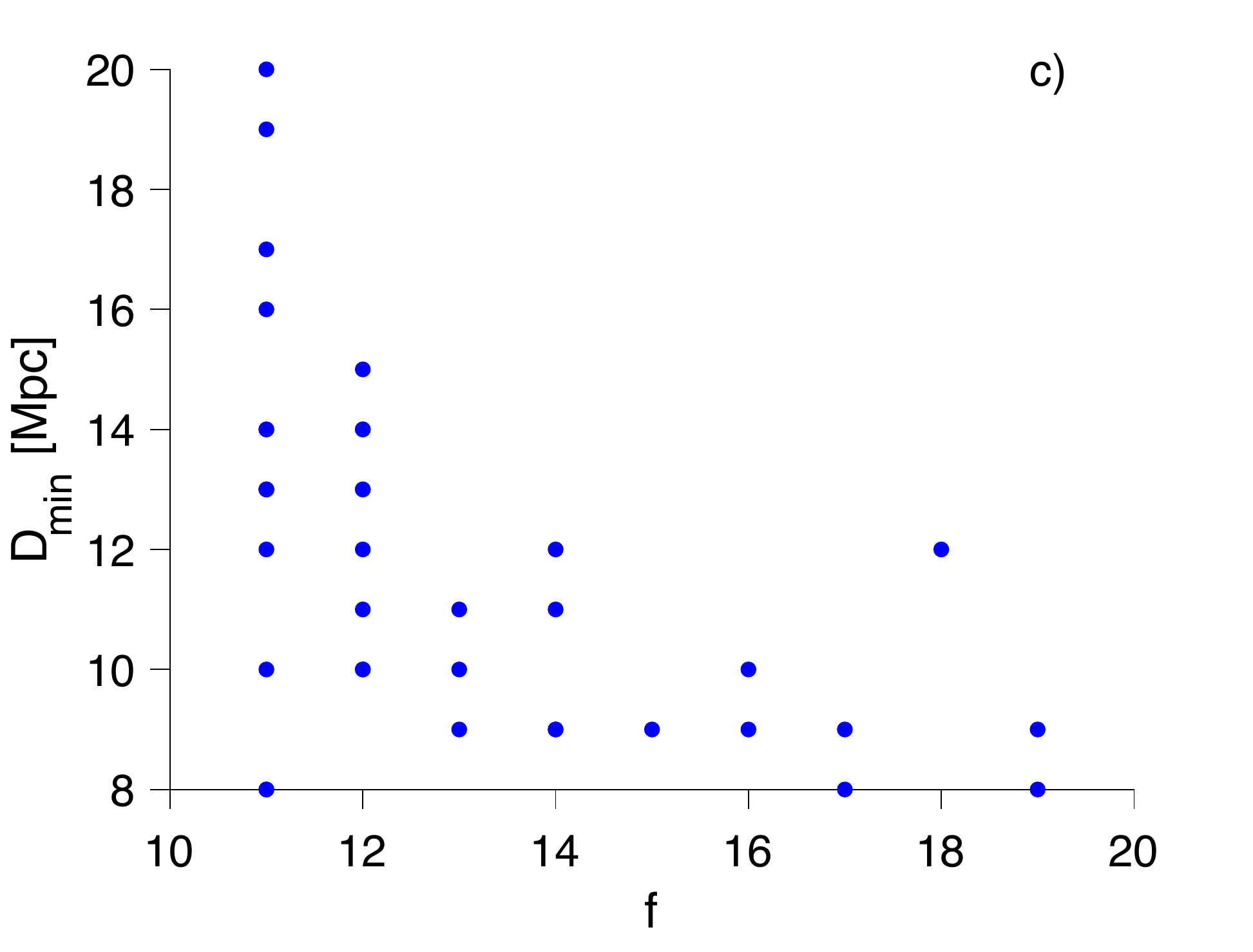}
      \caption{\footnotesize a) Distribution of best $f$ values with
redshift, $z$, for the 42 supercluster volumes of our sample.
b) Distribution of best linking lengths, $D_{min}$, with $z$. 
c) Linking length \textit{vs} segmentation parameters.}
      \label{optmz}
\end{figure}

A filament is consider detected if it has three or more nodes connected 
and the mean number density of its member galaxies is higher than the box 
volume number density $d_{bas}$.
In addition, we consider an optimal detection when, by increasing the 
linking length, the number of detected filaments is maximized (that is, 
the largest number of filaments before they begin to percolate). 
The combination of these two conditions in the parameter test space 
allows to find the best values for both $f$ and $D_{min}$. 
The results of the optimization process for the supercluster MSCC-454
are shown in Fig. \ref{Fil_test}. 
For this supercluster, the optimal parameter configuration is $f=10$ and 
$D_{min}=6$, with 9 filaments detected, as can be seen in panel a.
Fig. \ref{Fil_test}b depicts the projected distribution of the found 
filaments for the optimal parameter configuration, while 
Fig. \ref{Fil_test}c depicts the results for the second optimal 
configuration, $f=20$ and $D_{min}=8$.
In Fig. \ref{optmz}a, we present the distribution of best $f$ values 
for the superclusters of our sample with their mean redshift.
We can see a correlation between these parameters, with values of $f$
decreasing with $z$ -- that is, indirectly with the galaxy density in
the box. $D_{min}$ (Fig. \ref{optmz}b), on the other hand, grows
with $z$, in way to compensate the decrease in $f$. The space of best
values for $f$ and $D_{min}$ is shown in Fig. \ref{optmz}c.

\quad
\bigskip

\noindent \textbf{5. Discussion and conclusions}
\smallskip    

We have developed a new strategy for identifying systems of galaxies
(GSyF algorithm), correcting the FoG effect and detecting filaments 
(GFiF algorithm) inside superclusters of galaxies.
We also probed the most important free parameters for the two algorithms:
the minimum density contrast for detecting the systems and filaments, 
the cut in the number of groups for HC method (segmentation parameter)
and the linking length for constructing the graphs from which the 
filaments are split.

We have shown, by using mock maps, that the GSyF algorithm has a completeness 
above 90\% for 40 of our superclusters (in a total of 42),
with a contamination around 10\%.

With the application of the developed methodology to the sample of 
superclusters described in the present work, we generated a sample of
galaxy systems (from rich clusters to moderately rich groups), for which
we identified the galaxy members ($N_{mem}$) and measured their $R_{vir}$ and  
$M_{vir}$. Moreover, we detected filaments of galaxies connecting some of these 
systems. With this results we are able to study the effects of the 
LSS environment over the evolution of the hosted galaxies.  

It is worth noting that the grouping of galaxies allows to reduce 
considerably the computational time (consider that a supercluster 
has tens of thousands of galaxies).

\smallskip

\quad

{\bf Acknowledgments} The authors thank CONACyT and DAIP-UG for the
finantial support to I.S.-B. during her stay in IRAP.




\newpage
\appendix

\quad
\bigskip

\noindent \textbf{Appendix A}
\smallskip    

\vspace{-0in}
\begin{table}[h!]
\small
   \caption*{\textbf{Algorithm 1: }Galaxy systems finding algorithm 
(GSyF)} \label{tab:alg1}
\smallskip
   \textbf{Input:} Supercluster's galaxy positions $\alpha,~\delta$ and $z$.\\
   \textbf{Output:} Galaxy systems, membership and member galaxy positions 
corrected for FoG effect.
   \begin{enumerate}[nosep]
\item  {Compute local surface density of galaxies using VT.}
\item  {Construct 1,000 randomizations of galaxy positions to calculate the 
baseline surface density, $d_{bas}$.}
\item  {Calculate the density contrast of galaxies, $\delta_i$, respect to
$d_{bas}$.}	
\item  {Select galaxies with density contrast above the reference value 
$(1 + g)~d_{bas}$.}
\item  {Group galaxies by their position ($\alpha,~\delta$ and $1,000\times z$), 
using HC algorithm.}
\item  {Filter resulting groups by number of members N$_i \geq 3$.} \newline
\noindent \textbf {Virial refinement and FoG correction}
	\begin{enumerate}[nosep] 
	\item \textit{Select galaxies in a cylinder of radius 
$R_a = 1~h_{70}^{-1}$~Mpc projected in the sky, centered on the brightest
galaxy close to the group's centroid, and of longitude of 
$\Delta z~\pm$~3000~km~s$^{-1}$ along the line-of-sight, centered on the 
group mean $z$.}
	\item \textit {Calculate virial radius using Eq. \ref{R_V} using the 
bi-weighted velocity dispersion, harmonic radius and redshift.}
	\item \textit {Update $R_a$ by $R_{vir}$, mean $z$ by $v_{LOS}$ and
$\Delta z$ by $3\times \sigma_v$}.
	\item  {Compute iteratively virial radius for each group until 
$R_a \rightarrow R_{vir}$.}
	\item  {Calculate $N_{mem}$ and $M_{vir}$.}
	\item  {Correct comoving distance of the member galaxies for FoG effect
by re-scaling the cylinder length to the $R_{vir}$ size.}
	\item  {Calculate galaxy corrected rectangular coordinates.}
	\end{enumerate}
   \end{enumerate}
\end{table}

\begin{table}[h!]
\small
   \caption*{\textbf{Algorithm 2: }Mock maps generator} \label{tab:alg2}
\smallskip
   \textbf{Input:} Supercluster volume and number of galaxies. \\
   \textbf{Output:} Simulated distribution of galaxies in clusters, 
groups and field of the supercluster.
   \begin{enumerate}[nosep]
\item {The simulated volume is filled with $ N_{synth}$ synthetic 
systems of galaxies with $N_{elem} = {10-200}$.
The number of systems in the volume is set using the power function 
$\log_{10}[N_{synth}(N_{elem})][h_{70}^3~Mpc^{-3}]=m\log_{10}(N_{elem})$ 
(multiplicity function) with a slope $m$ set according to:} 		
   \smallskip
      \begin{center}
			$
			m= 
			\begin{cases}
			-2.48\qquad {\rm if}\ z<0.08\\ 
			-2.72\qquad {\rm otherwise} 
			\end{cases}
			$			 		
   \smallskip
      \end{center}
\item Set mock systems center position randomly in the volume.
\item Calculate synthetic system proxies (Mass $M_{500}$, radius $R_{vir}$ 
and velocity dispersion  $\sigma_v$).
\begin{center}
$
\begin{cases}
   \log_{10}M_{500}=1.03\log_{10}(N_{elem}-2.63)+0.34 \\
   \log_{10} R_{vir}=1.05~\log_{10}(M_{500}-0.35)+8.48 \\
   \log_{10} \sigma_v^3=2.33~\log_{10}(M_{500}-0.21)+3.04
\end{cases}	
$
\end{center}
\item Fill the systems with galaxies (elements) following a normal 
distribution  $N_{elem}({\mu_i},R_{vir})$ with $\mu_i = \{\alpha_i, \delta_i\}$.
\item Add FoG effects to galaxy systems by adding velocity dispersion $N_{elem}(z_i, \sigma_v)$.
\item Add random galaxies to the box volume following the ratio: 
60\% galaxies are located in field, and 40\% in systems.
	\end{enumerate}
\end{table}

\end{document}